\documentclass[letterpaper]{JHEP}
\usepackage{epsfig}

\usepackage{amssymb}

\def\drawbox#1#2{\hrule height#2pt
        \hbox{\vrule width#2pt height#1pt \kern#1pt
              \vrule width#2pt}
              \hrule height#2pt}

\def\Fund#1#2{\vcenter{\vbox{\drawbox{#1}{#2}}}}
\def\Asym#1#2{\vcenter{\vbox{\drawbox{#1}{#2}
              \kern-#2pt       
              \drawbox{#1}{#2}}}}

\def\funda{\Fund{6.5}{0.4}}

\def\symm{\funda\kern-0.4pt\funda}

\def\Tr{\mathop{\rm Tr}}

\newcommand{\ket}[1]{|#1\rangle}
\newcommand{\bra}[1]{\langle#1|}
\newcommand{\be}{\begin{equation}}
\newcommand{\ee}{\end{equation}}
\newcommand{\bea}{\begin{eqnarray}}
\newcommand{\eea}{\end{eqnarray}}

\preprint{CALT-68-2338\\ CITUSC/01-026 \\ SLAC-PUB-8916 \\
{\tt hep-th/0107178}}

\title{The Stringy Quantum Hall Fluid}

\author{Oren Bergman and Yuji Okawa
\\
California Institute of Technology, Pasadena CA 91125, USA
\\ and\\
CIT/USC Center for Theoretical Physics \\
Univ. of Southern California, Los Angeles CA \\
\email{bergman@theory.caltech.edu, okawa@theory.caltech.edu}}
\author{John Brodie\\
Stanford Linear Accelerator Center\\
Stanford University\\
Stanford, CA 94305\\
\email{brodie@SLAC.Stanford.edu}}

\abstract{Using branes in massive Type IIA string theory,
and a novel decoupling limit, we provide an
explicit correspondence between non-commutative Chern-Simons theory
and the fractional quantum Hall fluid. 
The role of the
electrons is played by D-particles, the background magnetic field
corresponds to a RR 2-form flux, and the two-dimensional fluid is
described by non-commutative D2-branes.
The filling fraction is
given by the ratio of the number of D2-branes and the number
of D8-branes, and therefore by the ratio rank/level of the Chern-Simons 
gauge theory.
Quasiparticles and quasiholes are realized as
endpoints of fundamental strings on the D2-branes,
and are found to possess fractional D-particle charges and 
fractional statistics.}

\begin{document}

\baselineskip16pt
\parskip=4pt

\section{Introduction}

Many physical systems can be described,
at least in some dynamical regime, by gauge field theories.
Indeed, gauge field theories are considered fundamental 
in understanding the interactions of elementary particles.
In recent years, gauge theories have also been applied to the
study of low-energy systems in condensed matter physics
\cite{fradkin}, such as the quantum Hall fluid \cite{girvin}, 
and high-$\mbox{T}_c$ superconductors \cite{highTc}. In this case 
gauge theory provides an effective, rather than fundamental,
description, which still proves to be quite useful.
In both cases, a better understanding of the dynamics of
gauge field theories is clearly a worthwhile pursuit.

It has recently been conjectured by Susskind that
a two-dimensional quantum Hall fluid of charged particles with
filling fraction $\nu=1/k$ is described
by a non-commutative Chern-Simons (NCCS) gauge theory at level
$k$ \cite{susskind}.
This conjecture is based on an earlier realization that a 
two-dimensional charged fluid  
in a strong magnetic field is equivalent to a pure Chern-Simons (CS) 
gauge theory, where the gauge group is the group of area-preserving
diffeomorphisms (APD) \cite{susskind_bacal}. Susskind argued that
in order to fully realize the graininess of the fluid one must 
replace the above CS theory with a non-commutative $U(1)$ CS theory.
The APD CS theory can be thought of as the lowest order approximation
to the NCCS theory. Some evidence for this conjecture was provided 
in \cite{hellerman}, where a formal correspondence between
the Hilbert space of NCCS theory and Laughlin's wavefunctions
\cite{laughlin} was shown. (For a more detailed discussion of the
correspondence see \cite{karabali}).

Studies of string dualities have led to novel realizations
of various gauge field theories using configurations of branes,
and much has been learned about the dynamics of gauge theories from
these configurations \cite{GK_review}.
Of course it is not at all clear that nature chooses to
realize gauge field theories in this way. 
Nevertheless, it is useful to study a variety of realizations of gauge 
theories. In particular, certain realizations may provide
insights into a dynamical regime of the theory where other
realizations, perhaps even the true realization, are not as useful.
It is therefore an interesting question whether NCCS theory
can be realized on branes in string theory.

It is well known that D-branes in a constant $B$ field 
background, and in a certain zero-slope limit known as 
the Seiberg-Witten (SW) limit, give rise to non-commutative gauge 
theories \cite{CDS,seiberg_witten}.\footnote{See also a recent 
review \cite{douglas_nekrasov}.}
On the other hand, it is also known that the world-volume
theories of D-branes in massive Type IIA string theory \cite{Romans},
{\em i.e.} in the presence of D8-branes, contain CS terms
\cite{bergshoeff}. In particular, the theory on a D2-brane
near  D8-branes is three-dimensional Yang-Mills-Chern-Simons
(YMCS) theory, with some
scalars and fermions. 
It is therefore natural to look for a brane realization of NCCS theory
using D2-branes in massive IIA string theory with a $B$ field,
and this is indeed where we will find it.
Furthermore, we will show that a consistent massive IIA 
background with non-vanishing $B$ requires a non-vanishing
RR magnetic field $G_2$ as well. 
Since $B$ also induces a uniform D-particle
density in the D2-brane, the entire system is found to correspond
to a two-dimensional fluid (the D2-brane) of charged particles
(the D-particles) moving in a uniform magnetic field ($G_2$),
namely a quantum Hall fluid.
Thus the brane picture draws a correspondence between 
NCCS theory and the quantum Hall fluid.

Matrix theory \cite{bfss,ikkt} provides an alternative description
of non-commutative gauge theory in terms of lower-dimensional D-branes.
For example, a non-commutative configuration of two matrices $X^1,X^2$ 
in the matrix model corresponds to a D2-brane with a uniform density
of D-particles. Fluctuations about this configuration are then governed
by a non-commutative SYM theory \cite{seiberg}.
NCCS theory can also be derived from a certain matrix model
\cite{bbst,susskind,polychronakos}, where the quanta are identified 
with the electrons
of the quantum Hall fluid \cite{susskind}.
As we will show, this matrix model can also be realized in terms
of D-particles. In an appropriate decoupling limit, the dynamics 
of D-particles in massive IIA string theory with a $B$ field 
are governed by a pure CS matrix model, which in turn reproduces 
the three-dimensional NCCS theory on the D2-brane.

Strings play a central role in the D-particle matrix model.
Their presence on D-particles in massive IIA string theory
is a known consequence of charge conservation 
\cite{polchinski_strominger,bgl1}, which in the matrix model
formulation is encoded in the Gauss law constraint.
The latter exhibits a variety of solutions, including a 
non-commutative D2-brane with no strings.
Solutions containing both a D2-brane and strings ending on it 
are identified 
with quasiparticles and quasiholes in the quantum Hall fluid.
These are found to carry fractional D-particle charge, and
to exhibit fractional statistics.

The paper is organized as follows.
In section 2 we show that the world-volume theory
on D2-branes in massive IIA string theory with a constant $B$
field background reduces to pure NCCS theory in an appropriate
decoupling limit. This gives a derivation of the conjectured
correspondence of NCCS theory and the quantum Hall fluid.
In section 3 we describe the matrix model of D-particles in
massive IIA string theory, which gives rise to NCCS theory
on the D2-brane.
In section 4 we identify the quasiparticles and quasiholes of
the quantum Hall fluid, in both the spacetime picture
and the matrix model description, and determine their
charge and statistics.
Section 5 is devoted to our conclusions and outlook.

{\it Note added:}
While this paper was being finished,
we were informed of work by L.~Susskind and S.~Hellerman
on a string theory construction of the quantum Hall system
\cite{susskind_hellweman}.

\section{The non-commutative massive D2-brane as a QH fluid}

\subsection{D2-brane in massive IIA supergravity}

Our starting point is massive Type IIA supergravity
\cite{Romans}. In particular, we shall be interested in a
form of the (bosonic) action which includes the 9-form potential $C_9$
\cite{bergshoeff,Polchinski_vol_II}\footnote
{We use the notations of Section B.4 in \cite{Polchinski_vol_II}
for differential forms.}
\begin{eqnarray}
\label{massiveIIA}
 S &=& {1\over 2\kappa_{10}^2} \int d^{10}x\,
  \sqrt{-g}\left\{e^{-2\Phi}\Big[R + 4|d\Phi|^2
  -{1\over 2}|H|^2\Big] - {1\over 2}|\widetilde{G}_2|^2
  - {1\over 2}|\widetilde{G}_4|^2 -
   {1\over 2}M^2\right\} \nonumber\\[5pt]
 & - & {1\over 4\kappa_{10}^2} \int \left\{
    \mbox{} 2MdC_9 +
   B \wedge \left(dC_3+{1\over 2}MB\wedge B\right)
     \wedge \left(dC_3+{1\over 2}MB\wedge B\right)\right\} ,
\end{eqnarray}
where $\kappa_{10}^2 = (2\pi)^7(\alpha')^4/2$, and
the gauge-invariant field strengths are given by
\begin{eqnarray}
 H &=& dB, \nonumber \\
 \widetilde{G}_2 &=& G_2 - MB = dC_1 - MB, \\
 \widetilde{G}_4 &=& dC_3 - C_1\wedge H + {1\over 2}MB\wedge B
 \;. \nonumber
\end{eqnarray}
The scalar $M$ is an auxiliary field, and the above action reduces
to ordinary (massless) Type IIA supergravity when $M=0$.
The equation of motion for $C_9$ implies that $M$ must be (piecewise)
constant. The equation from varying $M$, with $B=0$, is 
\be
 M = \raisebox{1pt}{*}dC_9 \;,
\ee
so this field is sometimes referred to as the RR scalar field strength.
This suggests that 8-branes create discontinuities in the value of
$M$. In fact, the configuration corresponding to an 8-brane located at 
$x^9=0$ is given by \cite{bergshoeff}
\be
\label{sugra_solution}
\begin{array}{rcl}
 ds^2 &=& f^{-1/2}(x^9)dx^\mu dx^\nu\eta_{\mu\nu} + f^{1/2}(x^9)(dx^9)^2 \\
 e^{-4\Phi(x^9)} &=& f^5(x^9) \\
 M(x^9) &=& \mp f^\prime(x^9) \;,
\end{array}
\ee
where
\be
 f(x^9) = c - m|x^9| \;.
\ee
Here $c$ and $m$ are positive constants.
The 8-brane is a domain wall across which
the value of $M$ jumps from $-m$ to $m$.
The sign in the solution for $M$ is correlated with
the chirality of the unbroken supersymmetry, and therefore 
differentiates an 8-brane from an anti-8-brane.
In string theory the jump is quantized
in units of the D8-brane charge $\mu_8$,\footnote{The charge of a D$p$-brane
is given by $\mu_p = (2\pi)^{-p}(\alpha')^{-(p+1)/2}$ 
\cite{Polchinski_vol_II}.}
so for $2k$ D8-branes we get
\be
\label{M}
 M(x^9) = 2k \kappa_{10}^2 \mu_8 \epsilon(x^9) 
   = {k\epsilon(x^9)\over 2\pi\sqrt{\alpha'}}  \;,
\ee
where $\epsilon (x^9)$ is the unit step function.

Note that the above solution is ill-defined globally, since the dilaton,
and therefore the string coupling $g_s=e^\Phi$, blows up at a
finite distance from
the 8-brane.\footnote{This was not the case for
the solution originally written in \cite{bergshoeff}, which corresponds
to $m<0$ in our conventions. The latter presumably describes a combination
of an orientifold 8-plane and seven D8-branes (plus images),
which carries the same total charge as a single (anti) D8-brane.}
However, as long as we stay close enough to the 8-brane the background
is locally consistent.\footnote{A completely consistent
configuration of 8-brane requires the introduction of orientifold
planes (as in Type IA string theory) to cut off the growth of the dilaton.}
In particular this is the case in the decoupling limit
we will consider.

Now consider a D2-brane along $(x^1,x^2)$, and at $x^9>0$,
in the background of $2k$ D8-branes.
Supersymmetry is completely broken since there are six coordinates
with mixed (ND) boundary conditions.
The low-energy world-volume field theory of the D2-brane contains the usual
$D=3$ ${\cal N}=8$ vector multiplet, {\em i.e.} a gauge field, 
seven adjoint scalars $X^I$ ($I=3,\ldots,9$),
and eight adjoint fermions,
as well as $2k$ massive fundamental fermions $\chi_a$ 
($a=1,\ldots,2k$) from the
D2-D8 strings. The action takes the form
\be
 S_{D2} = S_{SYM} + S_{top} + S_{\chi} \;,
\ee
where
\be
 S_{SYM} = -{1\over g_{YM_3}^2}\int d^3\sigma\,\sqrt{-g} \left[
  \frac{1}{4} g^{\alpha\alpha'}g^{\beta\beta'}
              F_{\alpha\beta}F_{\alpha'\beta'}  + 
  {1\over 2 (2\pi\alpha')^2}g^{\alpha\beta}g_{IJ}
  \partial_\alpha X^I\partial_\beta X^J + \cdots  \right] \;,
\ee
\be
 S_{top} = \mu_2 \int \left[C_3 + C_1\wedge (2\pi\alpha' F+B)\right] \;,
\ee
and
\be
 S_{\chi} = \int d^3\sigma\,
  \overline{\chi}_a\left(iD\!\!\!\!/\,\, + 
             {X^9\over 2\pi\alpha'} \right) \chi_a \;,
\ee
and the YM coupling is given by 
$g_{YM_3}^2 = g_s/\sqrt{\alpha'}$\,.\footnote{For a D$p$-brane the YM
coupling is $g_{YM_{p+1}}^2 = g_s (2\pi)^{p-2} (\alpha')^{(p-3)/2}$.}

We now integrate out the fundamental fermions.
This has two effects:
it produces a Chern-Simons term for the gauge field via the parity anomaly
\cite{witten_gaume,redlich},
\be
 S_{CS} = {k\over 4\pi} \int d^3\sigma\, \epsilon^{\alpha\beta\gamma}
  A_\alpha\partial_\beta A_\gamma \;,
\ee
where $\epsilon^{012} = -\epsilon_{012} = -1$,
and a potential for $X^9$.
The precise form of 
the potential is not important for our purpose.

\subsection{Non-commutative D2-brane}

It is well-known that turning on a constant NSNS field, and taking
an appropriate decoupling limit, deforms an ordinary
D-brane world-volume gauge theory into a non-commutative
gauge theory \cite{seiberg_witten}.
In our case we consider a constant NSNS field along $(x^1,x^2)$,
$B_{12}=-B_{21}=B$,
which deforms the YMCS theory on the
D2-brane to a non-commutative YMCS theory.
In particular, the gauge part of the action becomes
\begin{eqnarray}
 S_{gauge} & = & -{1\over 4g_{YM_3}'^2}\int d^3\sigma \sqrt{-\hat{G}}
              \hat{G}^{\alpha\alpha'} \hat{G}^{\beta\beta'}
              \hat{F}_{\alpha\beta}*\hat{F}_{\alpha'\beta'} 
    \nonumber \\[5pt]
 && \mbox{} + {k\over 4\pi}\int d^3\sigma 
   \epsilon^{\alpha\beta\gamma}\left(
 \hat{A}_\alpha * \partial_\beta \hat{A}_\gamma
 + {2i\over 3}\hat{A}_\alpha * \hat{A}_\beta * \hat{A}_\gamma\right)\;,
\end{eqnarray}
where $\hat{G}^{\alpha\beta}$ is the open string metric given
in the Seiberg-Witten limit by \footnote{Our
convention for $B$ differs from \cite{seiberg_witten} in
that $B=2\pi\alpha'B_{SW}$, so in the decoupling limit $B\sim\alpha'$.}
\be
 \hat{G}^{\alpha\beta} = \left\{
 \begin{array}{ll}
  -(B^{-1}gB^{-1})^{\alpha\beta} & \;\; \alpha,\beta=1,2 \\
  g^{\alpha\beta} & \;\; \mbox{otherwise} \;,
 \end{array} \right.
\ee
and $g'_{YM_3}$ is the effective YM coupling defined by
\be
 g'^2_{YM_3} = g^2_{YM_3}\sqrt{\mbox{det}(Bg^{-1})} \;.
\ee
The star product is defined in terms of the non-commutativity
parameter \footnote{Our convention for the star product is
$$
f(x) * g(x)
= \left.
\exp \left( -\frac{i}{2} \theta^{\alpha \beta}
\frac{\partial}{\partial \xi^\alpha}
\frac{\partial}{\partial \zeta^\beta} \right)
f(x + \xi) g(x + \zeta) \right|_{\xi = \zeta = 0}.
$$
The ordering of the open string space-time fields corresponding 
to the vertex operators $A(X(\tau))$ and $B(X(\tau'))$ is taken
to be $A(X) B(X)$ when  $\tau < \tau'$.
This is the convention taken in \cite{okawa_ooguri1,okawa_ooguri2}, and
is different from the one in \cite{seiberg_witten}.}
\be
\label{theta}
 \theta^{\alpha \beta} = 2\pi\alpha' (B^{-1})^{\alpha \beta} \;.
\ee
This is most easily understood by turning on the $B$ field before
integrating out the fundamental fermions.
The resulting non-commutative YM theory
with fundamental (Dirac) fermions exhibits a parity anomaly analogous
to the commutative case, and a one-loop non-commutative Chern-Simons
term is produced \cite{chu}. 

In the Seiberg-Witten zero-slope limit the theory contains both 
a non-commutative
YM term and a non-commutative CS term, as well as scalar fields and 
fermions. We shall instead consider a slight modification of this limit,
in which
\be
\label{newlimit}
\begin{array}{lcll}
 g_{\alpha\beta} &\sim & (\alpha')^{2+\delta} & 
   \quad (\alpha,\beta=1,2) \\[5pt]
 g_s &\sim & (\alpha')^{3/2+\epsilon} & ,
\end{array}
\ee
where $\delta > \epsilon > 0$. The Seiberg-Witten limit
corresponds to the case that $\delta = \epsilon =0$,
which keeps $\theta^{\alpha \beta}$, $\hat{G}^{\alpha \beta}$
and $g_{YM_3}'$ finite.
In the new limit $\theta^{\alpha \beta}$ remains finite,
but $\hat{G}^{\alpha \beta}$ and $g_{YM_3}'$ scale as
\begin{equation}
g_{YM_3}'^2 \sim (\alpha')^{\epsilon-\delta}, \qquad
\hat{G}^{\alpha\beta} \sim (\alpha')^{\delta}
\quad (\alpha,\beta=1,2) \;.
\end{equation}
In this limit the YM term vanishes, since
\be
 \frac{1}{2 g_{YM_3}'^2} \sqrt{-\hat{G}}
 \hat{G}^{11} \hat{G}^{22} \hat{F}_{12} * \hat{F}_{12} \sim
 (\alpha')^{2\delta-\epsilon}
\ee
and
\be
 \frac{1}{2 g_{YM_3}'^2} \sqrt{-\hat{G}}
 \hat{G}^{00} \hat{G}^{\alpha \alpha}
 \partial_t \hat{A}_\alpha * \partial_t \hat{A}_\alpha \sim
 (\alpha')^{\delta-\epsilon} \;.
\ee
In addition the scalar fields decouple, since the effective coupling 
constant for the canonically normalized fields
\begin{equation}
 \phi^I = \frac{(-\hat{G})^{1/4}}{g_{YM_3}'}
\frac{\sqrt{g_{II}}}{2 \pi \alpha'} X^I \;,
\end{equation}
is given by $g_{YM_3}'/(-\hat{G})^{1/4}\sim (\alpha')^{\epsilon/2}$.
Thus any interaction term written in terms of
$\hat{A}_\alpha$ and $\phi^I$ vanishes, because
this effective coupling constant vanishes,
and $\hat{G}^{\alpha \beta}$, which contracts indices from
the non-commutative gauge field, vanishes.
It is straightforward
to extend the argument to include the fermions.
This leaves us with the pure non-commutative Chern-Simons
(NCCS) theory,
\be
S_{NCCS} = {k\over 4\pi}\int d^3\sigma 
   \epsilon^{\alpha\beta\gamma}\left(
 \hat{A}_\alpha * \partial_\beta \hat{A}_\gamma
 + {2i\over 3}\hat{A}_\alpha * \hat{A}_\beta * \hat{A}_\gamma\right) \;.
\ee

Note also that the limit in (\ref{newlimit}) also implies
that in the supergravity 
solution (\ref{sugra_solution}) we must take 
$c\sim (\alpha')^{-6/5 -4 \epsilon /5}$.
In other words, the point where the dilaton diverges is taken
to infinity, and $g_s$ goes to zero at the location of the
D2-brane. 
The supergravity soultion is therefore well-defined.

\subsection{Quantum Hall fluid}

There is an apparent problem with the previous discussion.
Unlike massless supergravity, the equations of motion of massive
supergravity (\ref{massiveIIA}) do not admit a solution with a
nontrivial constant $B$ field and all other fields trivial.
Another way to see this is that the $B$ field induces sources
for the RR fields inside the world-volume of the 8-branes.
In the simplest case of a single non-trivial constant component 
$B_{12}=B$,
and a trivial $C_3$, the equation of motion for the NSNS field is given by
\be
 d(e^{-2\Phi}\,\raisebox{1pt}{*}H) =
    M\,\raisebox{1pt}{*}(G_2 - MB) \;.
\ee
Since $H$ vanishes we find that the solution requires
a $G_2$ flux along $(x^1,x^2)$
\be
\label{G2flux}
 G_2 = MB = {kB\over 2\pi\sqrt{\alpha'}}\epsilon(x^9) \;.
\ee
This can also be understood from the fact that
the constant $B$ field 
induces a uniform D6-brane charge density per unit area in the
$(x^1,x^2)$ plane in the world-volume of the
D8-brane. From the coupling
\be
 \mu_8\int C_7\wedge B 
\ee
in the D8-brane world-volume theory, we deduce that
the D6-brane charge density for $2k$ D8-branes is
\be
 \rho_6 = 2k\mu_8 B =
  {2kB\over (2\pi)^8(\alpha')^{9/2}} \;.
\ee
The resulting magnetic field is given by
\be
 G_2 = {1\over 2}(2\kappa_{10}^2) \rho_6 \epsilon(x^9)
     = {kB\over 2\pi\sqrt{\alpha'}}\epsilon(x^9) \;,
\ee
in agreement with (\ref{G2flux}).
We will assume that this relation continues to hold in the
presence of the D2-brane.

This is where the problem becomes a virtue.
The $B$ field induces a uniform D-particle charge density per
unit area in the membrane given by
\be
\label{0brane_density}
 \rho_0 = \mu_2 B = { B \over (2\pi)^2(\alpha')^{3/2}} \;.
\ee
The non-commutative membrane can therefore be thought of
as a two-dimensional fluid of RR-charged D-particles in a background RR
magnetic field, {\em i.e.} a quantum Hall fluid.
The filling fraction $\nu$ is defined as the ratio of the carrier
density to the degeneracy of the Landau level. The latter is given by
$\mu_0 G_2/(2\pi)$, so the filling fraction is
\be
\label{filling_fraction}
 \nu = {\rho_0/\mu_0\over \mu_0|G_2|/(2\pi)} = {1\over k} \;.
\ee
This establishes Susskind's conjecture that a quantum Hall fluid
at filling fraction $1/k$ is described by a NCCS theory
at level $k$ \cite{susskind}.
In this approach it is clear why the inverse filling fraction
is quantized; it simply corresponds to the quantized value of
the RR scalar field strength $M$, or equivalently to (half) the
number of D8-branes. It is less clear at this stage whether the 
D-particles in the membrane behave like 
fermions or bosons, and how this property correlates with $k$.
This will be clarified in the matrix model picture in the next section.
\begin{figure}
\centerline{\epsfxsize=4in\epsfbox{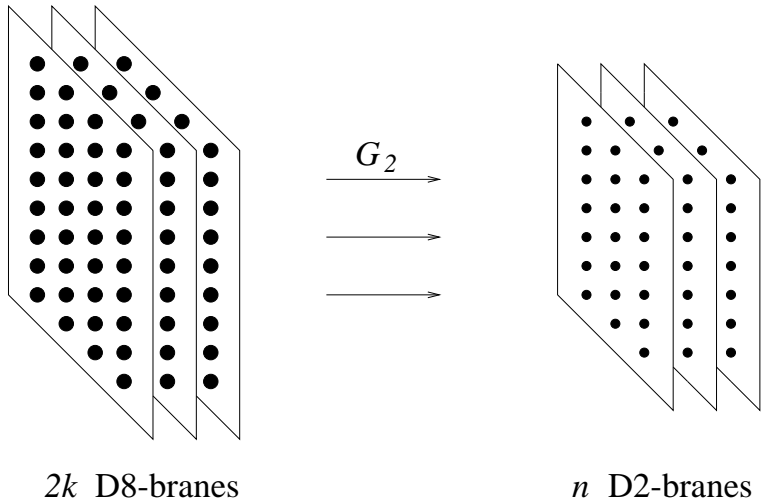}}
\hspace{1cm}
\medskip
\caption{Brane configuration of the quantum Hall fluid with $\nu=n/k$.
A uniform $B$ field induces D-particles (small circles) inside the
D2-branes, and D6-branes (large circles) inside the D8-branes.
The D6-branes give rise to a uniform magnetic RR field $G_2$.}
\end{figure}

\subsection{Multiple D2-branes}

It was also conjectured that the theory for a  quantum Hall fluid with a 
filling
fraction $\nu=n/k$ would be a level $k$ $U(n)$ NCCS theory
\cite{susskind}. In our construction this is a trivial
generalization of $\nu=1/k$ case. Consider $n$ D2-branes
in massive supergravity with $2k$ D8-branes.
The resulting world-volume theory is $U(n)$ YMCS theory with
adjoint matter, and a CS coefficient $k$.
In the presence of the $B$ field, and in our scaling limit,
this becomes a level $k$ $U(n)$ NCCS theory.
The RR magnetic field is still given by (\ref{G2flux}), but the
D-particle charge density is now
\be
 \rho_0 = n\mu_2 B \;,
\ee
so the filling fraction is
\be
 \nu = {n\over k}\;.
\ee
The challenge is to understand, in this picture, which fractions
actually exhibit a mass-gap, and why there is a hierarchy of fractions.

\section{D-particles and the Chern-Simons Matrix Model}

It is well-known that non-commutative Yang-Mills theory
on a D$p$-brane with $2n$ non-commutative directions
can be obtained from a non-commutative configuration
of D($p-2n$)-branes \cite{seiberg}. In particular the 
non-commutative SYM theory on the massless Type IIA D2-brane
is obtained by considering fluctuations about a non-commutative
configuration in the D-particle matrix model.
Similarly, NCCS theory can also be obtained from a one-dimensional 
matrix model \cite{bbst}
\be
S = \int dt~ {\rm Tr} \left[
-kA + \frac{k}{2} (\theta^{-1})_{ij} X^i D_0 X^j \right] \;,
\label{NCCS-matrix-model}
\ee
expanded about a non-commutative background
\begin{equation}
\label{fluctuations}
  X^i = x^i + \theta^{ij} \hat{A}_j
\end{equation}
where
\begin{equation}
[x^i, x^j] = -i \theta^{ij}.
\end{equation}
To make this consistent for finite $N$, it was proposed in 
\cite{polychronakos} to add a term
\be
\label{boson}
 i\int dt \psi^\dagger D_0 \psi \;,
\ee
where $\psi$ is a complex boson transforming in the fundamental
representation of $U(N)$.
We will show that this matrix model can be realized
by D-particles in massive IIA supergravity with a 
$B$-field in the scaling limit described in the previous section.

\subsection{D-particles in massive IIA supergravity}

It has been known for quite some time that there are subtleties
involved in introducing D-particles in massive IIA supergravity,
{\em i.e.} in the presence of D8-branes \cite{polchinski_strominger,bgl1}.
In particular, charge conservation requires strings to end on 
D-particles, whereas this is forbidden in massless IIA \cite{strominger}.
In the spacetime picture this follows from the equation of motion
for $B$, which for trivial background fields becomes
\be
 d(e^{-2\Phi}\,\raisebox{1pt}{*}H) =
    M\raisebox{1pt}{*}G_2 + 
  {2\kappa_{10}^2\over 2\pi\alpha'} n_s \delta_8 \;,
\ee
where the second term corresponds to $n_s$ fundamental string sources.
Integrating over an 8-sphere surrounding $N$ D-particles we find
that 
\be
 n_s = - 2\pi\alpha' M N \mu_0 = - kN\epsilon(x^9) \;,
\ee
where we have used the value of $M$ for $2k$ D8-branes
at $x^9=0$ (\ref{M}). It follows that 
$k$ strings are attached to each D-particle,
and their orientation depends on which side of the D8-branes
the D-particle is.
Alternatively, if a string ends on any other D-brane,
the above equation implies that its end carries $1/k$ units of
D-particle charge. This fact will be important in the next section.

Another way to see that strings are attached to the D-particle
is via the world-line theory \cite{dfk}.
In this picture the end of the string 
is a source for the world-line electric field, and is necessary
in order to cancel an induced electric charge due to the
background D8-branes. 
In the usual decoupling limit
\be
 \alpha'\rightarrow 0\;,\; g_{ij}\sim (\alpha')^2 \;,\;
 g_s \sim (\alpha')^{3/2} ,\;\;
 \mbox{everything else}\sim O(1) \;,
\ee
the world-line theory for $N$ D-particles is given by \cite{bgl1}
\begin{eqnarray}
\label{08matrix_model}
 S  &=&  {1\over g_{YM_1}^2}\int dt\,\mbox{Tr}\, \bigg[
 {1\over 2(2\pi\alpha')^2} g_{ij} D_0X^iD_0X^j + 
 {1\over 4(2\pi\alpha')^4} g_{ik}g_{jl}[X^i,X^j][X^k,X^l] + \cdots\bigg]
  \nonumber\\[5pt]
  && \mbox{} + \int dt\,\left\{-\mbox{Tr}\bigg[kA_0 + 
   {kX^9\over 2\pi\alpha'}\bigg] +
     \psi^\dagger_a \Big(iD_0 - {X^9\over 2\pi\alpha'}\Big)\psi_a 
     \right\}\;,
\end{eqnarray}
where $\psi_a$ ($a=1,\dots,2k$) are $N$-component vectors of
fermions coming from the (R sector of the) 0-8 strings,
and we have suppressed the terms involving the fermionic partners
of $X^i$. 
Each of the fundamental fermions contributes a CS term
with coefficient $-1/2$, as well as a 
linear $X^9$ potential,  to the one-loop effective action \cite{bss,bgl1}.
The former corresponds to an induced charge, and the latter to
a non-vanishing repulsive force. In order to cancel these we must
add a bare CS term and an attractive potential as above. 
The former is interpreted as the charge at the end of the string, 
and the latter is due to the tension of the string.

\subsection{The Chern-Simons matrix model}

Now consider turning on a constant $B$ field along $(x^1,x^2)$.
As we have seen in the previous section, this requires turning
on a constant $G_2$ along $(x^1,x^2)$ as well.
This introduces additional terms in the equation of motion for $B$,
\begin{eqnarray}
 d(e^{-2\Phi}\,\raisebox{1pt}{*}H) &=&
    M\raisebox{1pt}{*}G_2 - M^2 \raisebox{1pt}{*}B
   - M \raisebox{1pt}{*}dC_3\wedge B
    \nonumber \\
 && \mbox{}  + d \left( \raisebox{1pt}{*} dC_3 \wedge C_1
               -\frac{1}{2} C_3 \wedge d C_3 \right) 
    + {2\kappa_{10}^2\over 2\pi\alpha'} n_s\delta_8 \;.
\label{B_equation}
\end{eqnarray}
In particular, the third term contributes to the integral
over an eight-sphere at infinity if D2-branes are present,
and one gets
\be
\label{strings1}
 n_s = \mbox{} -kN + {k B \over (2\pi)^2\alpha'} nA 
 = \mbox{} -kN + {G_2 \over 2\pi\sqrt{\alpha'}} nA \;,
\ee
where we have used (\ref{G2flux}) in the second equality.
Here $n$ is the number of 
D2-branes, and $A$ is their (infinite) area.
Using the definition of the filling fraction $\nu$ 
(\ref{filling_fraction}), we can rewrite this condition as
\be 
\label{strings}
 n_s = \mbox{} -N\left(k - {n\over \nu}\right) \;.
\ee
If $n=0$ the situation is the same as before,
and each D-particle has $k$ strings attached.
For $n\neq 0$ the number of strings depends on the filling fraction.
In particular for $\nu = n/k$ we find $n_s=0$. Thus no strings are 
attached to the D-particles in the quantum Hall fluid phase.
In the next section we will consider a more general situation
in which the filling fraction is different from this value,
which requires strings to end on the D2-branes.

The world-line theory is also modified in the presence of 
$B$ and $G_2$.
First, there will be an additional $X^9$ potential due to the
induced D6-branes. More importantly though, there is a Lorentz-type 
interaction with the background RR field \cite{taylor1,taylor2,myers}
\be
\label{G_2coupling}
 S_{G_2} = {\mu_0\over 2}\int dt\, (G_2)_{ij} \mbox{Tr} 
   \left[X^iD_0 X^j\right] \;.
\ee
Now consider the decoupling limit (\ref{newlimit}),
in which $g_{YM_1}^2 = g_s (2\pi)^{-2}(\alpha')^{-3/2} \sim 
(\alpha')^\epsilon$.
In this limit the fields $X^I$, with $I=3,\ldots,9$, and 
the adjoint fermions decouple completely. In addition, 
the CS term (\ref{G_2coupling}) dominates over the kinetic terms
for $X^1$ and $X^2$, and the only non-trivial part of the matrix model
is given by
\begin{eqnarray}
\label{CSmatrix}
 S &=& \int dt\Big\{ \mbox{Tr} 
   \left[{1\over 2\sqrt{\alpha'}}(G_2)_{ij} X^i D_0 X^j - kA_0 \right] + 
  i \psi^\dagger_a D_0 \psi_a \Big\} \nonumber \\[5pt]
 &=& \int dt\Big\{ \mbox{Tr} 
   \left[{k\over 2}(\theta^{-1})_{ij} X^i D_0 X^j - kA_0 \right] + 
  i \psi^\dagger_a D_0 \psi_a \Big\} \;,
\end{eqnarray}
where we have used (\ref{G2flux}) and (\ref{theta}) in the
second equality.
This is almost identical to the finite $N$ CS matrix model 
discussed in \cite{polychronakos}. The only difference is that here 
the fields $\psi_a$ are fermions, whereas in \cite{polychronakos}
they are treated as bosons. The former is more natural, given 
the form of their action. Recall that these fields correspond 
to 0-8 strings.

\subsection{Solutions}

The dynamics of the matrix model in (\ref{CSmatrix}) is 
determined completely from the Gauss law constraint,
\be
\label{EOMmatrix}
 -i(\alpha')^{-1/2}G_2[X^1,X^2]_{mn} + 
  \psi_{a,m}^{\phantom{\dagger}}\psi_{a,n}^\dagger = 
  k\delta_{mn} \;.
\ee
Note that for $G_2=0$ we get
\be
 \psi_{a,m}^{\phantom{\dagger}}\psi_{a,n}^\dagger = k\delta_{mn} \;.
\ee
This is just the statement that $k$ strings are attached to each
D-particle.

For finite $N$ the trace of the constraint is always
\be
 \mbox{Tr}\,\psi^{\phantom{\dagger}}_a\psi^\dagger_a = Nk \;,
\ee
since the trace of the commutator vanishes.
However, when $N\rightarrow\infty$ more general solutions to the
trace constraint are possible.
In particular, 
the quantum Hall fluid corresponds to a non-commutative D2-brane,
which is described by a D-particle configuration $X^1=x^1$,
$X^2=x^2$, and $X^I=0$ ($I=3,\ldots,9$), in which
\be
 [x^1,x^2]_{mn} = -i\theta^{12}\delta_{mn} = i\theta\delta_{mn} \;,
\ee
where we have defined $\theta = -\theta^{12}$.
It then follows from (\ref{theta}) and (\ref{G2flux}) that
\be
 \psi_a^{\phantom{\dagger}}\psi_a^\dagger= 0 \;,
\ee
namely that there are no strings. This is consistent with what was found
in the spacetime viewpoint (\ref{strings}).

By considering fluctuations of $X^1$ and $X^2$ about this 
solution (\ref{fluctuations}), and replacing
\be
 \mbox{Tr} \rightarrow 
{{\rm Pf}(\theta^{-1})\over 2\pi}\int dx^1 dx^2
= {1 \over 2\pi\theta}\int dx^1 dx^2 \;,
\ee
one obtains NCCS theory \cite{susskind}.
The D-particle density per unit area in this configuration
is $1/(2\pi\theta)$, so the filling fraction is $1/k$,
in agreement with the result in Section 2.

The advantage of the matrix model approach is that it allows us to 
address the question of the statistics of the D-particles, {\em i.e.}
the electrons of the QH fluid. In general, D-particles have rather
complicated ``non-abelian'' statistics, due to the fact that their
positions are described by matrices rather than numbers.
However, in the non-commutative configuration described above
the effective two-dimensional statistics in the $(x^1,x^2)$ plane 
simplifies. As shown in \cite{susskind}, the phase associated with
exchanging two D-particles in the D2-brane configuration is given by
$\exp(i \pi k)$. This was used in \cite{susskind} to argue that $k$
should be quantized. In our case $k$ is (half) the number 
of D8-branes, so it is manifestly quantized. We therefore see that
the D-particles behave as bosons if $k$ is even, and fermions 
if $k$ is odd.

\section{Quasiparticles and quasiholes}

The quantum Hall fluid at filling fraction $1/k$ (with $k$ odd) 
is a stable state with a mass-gap \cite{laughlin}. 
The lowest-lying excitations of this state
correspond to adding or subtracting a unit of magnetic flux,
and thereby slightly shifting the filling fraction.
The former is known as a {\em quasihole}, and the latter as
a {\em quasiparticle}. Due to the incompressibility of the quantum
Hall fluid, it can be shown that these excitations effectively
carry a fractional charge $\mp 1/k$, and therefore exhibit fractional
statistics. The total charge of the fluid
remains the same, but the additional flux causes a redistribution
of the charge density.
In this section we will identify the quasiparticles and
quasiholes in our string picture, and show that they have
the correct charge and statistics.

\subsection{Space-time picture}

Let us begin with a single D2-brane, and a filling fraction
$\nu=1/k$. Now consider a shift in the RR magnetic field
corresponding to one flux quantum, {\em i.e.}
\be
 G_2 \rightarrow G_2 + \Delta G_2 \;,
\ee
where $\Delta G_2 = \pm 2\pi\sqrt{\alpha'}/A$. 
This can be achieved by creating a D6-$\overline{\mbox{D6}}$
pair, and moving the D6 (in the + case) or the $\overline{\mbox{D6}}$
(in the - case) across the D2-brane (figure~2).
Since the total D-particle
number $N$ is kept fixed, this changes the filling fraction to
\be
 \nu' = {1\over k \pm {1\over N}} \;.
\ee
The condition (\ref{strings}) now requires $n_s=\pm 1$.
This can also be understood from the fact that when the D6
(or $\overline{\mbox{D6}}$) crosses the D2-brane a string
is created between them. We therefore identify the quasiparticle
(quasihole) with a string ending on the D2-brane.
This also explains, from the spacetime point of view, why 
quasiparticles and quasiholes correspond to sources in the
world-volume NCCS theory \cite{susskind}.

The above result suggests that the quasiparticle and quasihole
carry a fractional D-particle number $\pm 1/k$, since a whole
D-particle (outside the D2-brane) requires $k$ strings. 
Indeed, if $G_2=B=0$, the equation of motion for $B$ (\ref{B_equation})
implies that the end of a string on any D-brane in massive
supergravity carries $1/k$ units of D-particle charge.
In our case the background fields are non-trivial, and the total
D-particle charge is held fixed. The latter can be 
thought of as coming from two components
\be
 N = (N+\Delta N) - \Delta N \;,
\ee
where $\Delta N = \pm 1/k$. The first component corresponds to 
a quantum Hall fluid with filling fraction given by
\be
\nu = {2\pi\sqrt{\alpha'}\over A}{N + \Delta N\over G_2 + \Delta G_2} 
    = {1\over k}\;,
\ee
and the second component is the quasihole (or quasiparticle).
The idea is that when we shift $G_2$ 
the D-particle density changes to maintain the quantized
filling fraction. Conservation of total D-particle charge 
then requires a charge deficit (or excess) somewhere,
and this is provided by the end of the string.
The new state can be thought of as consisting of two components
\be
 \Psi = \Psi_{f} \cdot \Psi_{qp} \;,
\ee
where $\Psi_f$ describes the fluid at $\nu=1/k$, and $\Psi_{qp}$
describes the quasiparticle.
This will be made somewhat more quantitative in the matrix model approach.

Consider now a pair of quasiparticles corresponding to
two strings ending on the D2-brane at different positions. 
When we move one of the string ends around the 
other, the quasiparticle wavefunction acquires an Aharonov-Bohm
phase. 
Relative to the case with a single
quasiparticle, the phase given by
\be
 \phi = A \Delta G_2 \cdot { \mu_0 \over k} =  
   \pm {2\pi\over k} \;.
\ee
Thus the quasiparticles exhibit fractional statistics.
\begin{figure}
\centerline{\epsfxsize=4.5in\epsfbox{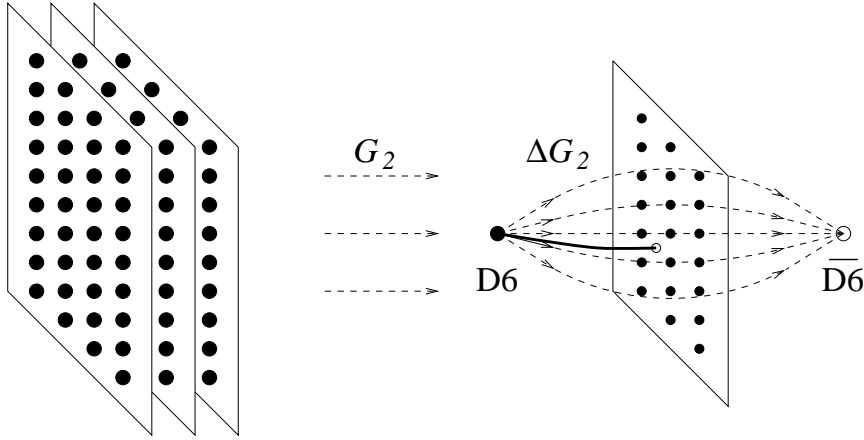}}
\hspace{1cm}
\medskip
\caption{Brane configuration of a quasihole. The D6-$\overline{\mbox{D6}}$
pair adds one unit of $G_2$ flux, as well as a D2-D6 string. The end of
the string on the D2-brane carries a deficit RR charge $\mu_0/k$,
represented by the unfilled circle.}
\end{figure}

\subsection{Matrix model picture}

The ground state of a quantum Hall fluid at filling fraction
$1/k$ corresponds to a solution of (\ref{EOMmatrix}) with
\begin{eqnarray}
 \psi^{\phantom{\dagger}}_{a} \psi^\dagger_{a} &=&  0 \\[5pt]
\label{NCmatrix} 
 \mbox{} [X^i,X^j] &=& -i\theta^{ij} \, {\bf 1} \;,
\end{eqnarray}
where 
\be 
\theta^{ij} = 2 \pi \alpha' (B^{-1})^{ij}
= \sqrt{\alpha'}k (G_2^{-1})^{ij} \;.
\ee
This corresponds to a non-commutative D2-brane with a uniform 
D-particle density
given by
\be
 {N\over A} = {(\theta^{-1})_{12} \over 2\pi} \;,
\ee
and no strings. 

A quasiparticle (or quasihole) corresponds to 
adding an additional unit of $G_2$ flux, while maintaining the
same filling fraction for the fluid component, and keeping the
total D-particle charge fixed. The latter condition requires
a string to end on the D2-brane, which in the matrix model 
corresponds to $\Tr\psi^{\phantom{\dagger}}_a \psi^\dagger_a
= \psi^{\phantom{\dagger}}_{a,n} \psi^\dagger_{a,n}
= \mp 1$. For example, if 
we choose 
\be
\label{psi}
 \psi^{\phantom{\dagger}}_a \psi^\dagger_a = \mp |0 \rangle \langle 0| \;,
\ee
Gauss' law is solved by
\be
\label{quasiparticle}
 [X^i, X^j] =  -i(\theta')^{ij}
 \left({\bf 1} \pm \frac{1}{k} |0 \rangle \langle 0| \right) \;,
\ee
where
\be
 (\theta')^{ij} = \sqrt{\alpha'}k
 \left( { 1 \over G_2 + \Delta G_2} \right)^{ij} \;.
\ee
This agrees with Susskind's prescription of introducing by hand a source
term to Gauss' law in order to describe a quasiparticle \cite{susskind}.
In our case the ``source'' is provided by the field $\psi$,
which fits with our identification of the quasiparticle and quasihole
as strings.
An explicit form of $X^i$ satisfying (\ref{quasiparticle})
was given in \cite{susskind}.

\subsection{Charge distribution}

Given any D-particle configuration $X^i$, one could in principle compute
the distribution of RR charges from the couplings of the D-particles
to the RR fields \cite{taylor1,taylor2,myers}.
In particular, the
(Fourier transform of) the D-particle density is
given by
\begin{equation}
J_{D0}(p) = {\rm Tr}\, e^{ipX} \;.
\end{equation}
This is derived from the coupling to $(C_1)_0$:
\begin{equation}
\mu_0 \int dt~{\rm Tr}~ (C_1)_0 (X)
= \mu_0 \int dt \int \frac{d^2 p}{(2 \pi)^2}
(\tilde{C}_1)_0 (-p) J_{D0}(p) \;,
\end{equation}
where
\begin{equation}
(\tilde{C}_1)_0 (p) = \int d^2 x~ e^{ipx} (C_1)_0 (x) \;.
\end{equation}
The total D-particle number is given by
$J_{D0} (0)= {\rm Tr}\,{\bf 1} = N$.
If we map the matrix configuration of $X^i$
to the non-commutative gauge field configuration $\hat{A}_i$,
the RR currents give the coupling of the non-commutative
gauge field to the RR potentials
\cite{okawa_ooguri2,mukhi_suryanarayana,liu_michelson}.
For example, $J_{D0} (p)$ is nothing
but the open Wilson line \cite{IIKK,DR,GHI}
\begin{equation}
\label{wilson_line}
J_{D0} (p) = \frac{{\rm Pf} (\theta^{-1})}{2 \pi} \int d^2 x~
\left[ e^{ipx}
P \exp \left( i \int_0^1 \hat{A}_i (x+l \tau) l^i d\tau \right)
\right] \;,
\end{equation}
where $l^i = p_j \theta^{ji}$ and the star product is implicit.
We can use the RR currents to study
the D-brane charge distribution
of non-commutative solitons as was done in 
\cite{hashimoto_ooguri}.\footnote{In the Seiberg-Witten limit,
there is no ambiguity
in the D-brane charge distribution
coming from field redefinition of RR potentials
when we set at least one transverse scalar to be zero.
The closed string on-shell condition can then be satisfied
for an arbitrary momentum in the noncommutative directions
\cite{okawa_ooguri1}.
We thank H. Ooguri for the discussion on this point.}
We would like to compute the $p$ dependence of $J_{D0}$,
and thereby the charge distribution, 
for the case of the quasiparticle (or quasihole)
(\ref{quasiparticle}).

It turns out to be technically challenging to compute $J_{D0}(p)$
using the explicit form of the matrices $X^1,X^2$ given by
\cite{susskind}. Instead, we will give an indirect argument.
It was shown in \cite{okawa_ooguri2,mukhi_suryanarayana,liu_michelson}
that the current $J^{0ij}(p)$
which couples to $(C_3)_{0ij}$
is topological, in the sense that its value does not change
for a small variation of the configuration $\delta X^i$.
This is physically reasonable because the D2-brane charge
does not change for a small fluctuation of the non-commutative
gauge field.
The explicit form of $J^{0ij}(p)$ is derived from the coupling
\begin{eqnarray}
-\frac{i}{2} \frac{\mu_0}{2 \pi \alpha'}
\int dt~ {\rm Tr}~ C_{0ij} (X) [X^i, X^j]
= \mu_2 \int dt \int \frac{d^2 p}{(2 \pi)^2}
\tilde{C}_{0ij} (-p) J^{0ij} (p),
\end{eqnarray}
and is given by
\begin{equation}
J^{0ij} (p) = -i \pi~ {\rm Tr}\, [X^i, X^j] e^{ipX} \;.
\end{equation}
For a configuration with a single D2-brane
\begin{equation}
J^{0ij} (p) = -i \pi~ {\rm Tr}\, [X^i, X^j] e^{ipX}
= \frac{1}{2} (2 \pi)^2 \epsilon^{ij} \delta (p) \;.
\end{equation}
For the quasiparticle solution (\ref{quasiparticle}) this gives
\begin{equation}
{\rm Tr}\, (\theta')^{ij}
\left( {\bf 1} \pm \frac{1}{k} |0 \rangle \langle 0| \right)
e^{ipX} = - 2\pi \epsilon^{ij} \delta (p).
\end{equation}
The D-particle density is therefore given by
\begin{equation}
J_{D0}(p) = {\rm Tr}\, e^{ipX}
= 2 \pi (\theta'^{-1})_{12} \delta (p)
\mp \frac{1}{k} \langle 0| e^{ipX} |0 \rangle.
\end{equation}
The first term corresponds to a uniform distribution
with density $1/(2\pi\theta')$, and 
the second term corresponds to the $1/k$ excess (or deficit) charge
carried by the quasiparticle (quasihole), which compensates
for the change in the uniform distribution from
$1/(2\pi\theta)$.
We can easily compute this term in two extreme limits.
For $k\rightarrow\infty$, the configuration reduces
to (\ref{NCmatrix}), and the matrices can be expressed
in terms of creation and annihilation operators
\be
 X^1 = \sqrt{\frac{\theta}{2}} (a + a^\dagger) \;, \;
 X^2 = -i \sqrt{\frac{\theta}{2}} (a - a^\dagger) \;,
\ee
We therefore find that
\begin{equation}
\langle 0| e^{ipX} |0 \rangle
= \exp \left[ - \frac{\theta}{4}(p_1^2 + p_2^2) \right] \;,
\end{equation}
and the coordinate space charge distribution is Gaussian.
For $k=1$, on the other hand, the computation reduces 
to the one done in \cite{hashimoto_ooguri},
and the momentum space charge distribution is given by
\begin{equation}
\langle 0| e^{ipX} |0 \rangle = 1 \;,
\end{equation}
so in coordinate space it is $\delta(x)$.
Since the charge deficit (or excess) is localized
in both limits, we expect that it is localized
in general.

The solution corresponding to a quasiparticle in (\ref{psi})
and (\ref{quasiparticle}) is not unique. For example, one can
choose 
$\psi^{\phantom{\dagger}}_a \psi^\dagger_a = |1 \rangle \langle 1|$, 
and similarly
replace (\ref{quasiparticle}). 
The precise form of the charge distribution of the quasiparticle
changes, but the total charge deficit (or excess) is still $1/k$.

Multi-quasiparticle (quasihole) states can be described by configurations
of the form 
\begin{eqnarray}
 \psi^{\phantom{\dagger}}_a \psi^\dagger_a &=& 
   \sum_{m=0}^N c_m \ket{m}\bra{m} \\
 \mbox{} [X^i, X^j] &=& -i (\theta')^{ij}
         \left({\bf 1} - \sum_{m=0}^N
         c_m \ket{m}\bra{m}\right) \;.
\end{eqnarray}
The total charge deficit (or excess) is simply given by
$\mbox{Tr}\, \psi^{\phantom{\dagger}}_a \psi^\dagger_a /k = \sum c_m/k$.
However the entire charge is concentrated at the origin.
It would be interesting to find multi-centered solutions.
In particular, one would like to identify
the collective coordinates of multi-quasiparticle states,
just as in the case of non-commutative solitons in 
\cite{gross_nekrasov}.
This would be particularly useful in verifying the statistical
properties of the quasiparticles in the matrix model,
as was done for the D-particles in \cite{susskind}.

\section{Conclusions and outlook}

In this paper we have derived the correspondence between
NCCS theory and the quantum Hall fluid. The brane picture
which we have presented, namely a D2-brane in massive IIA
string theory with a constant $B$ field background, in an
appropriate decoupling limit, can be viewed in two ways.
In the spacetime viewpoint the system is seen as a 
two-dimensional fluid of charged particles in a uniform background
magnetic field, where the charged particles are D-particles,
the magnetic field is the RR field $G_2$, 
and the fluid corresponds to a non-commutative configuration
of D-particles forming a D2-brane.
At the same time, the D2-brane world-volume theory 
is a pure NCCS gauge theory.

The brane picture naturally explains various aspects of the
quantum Hall fluid, such as the quantization of the filling fraction,
and the charge and statistics of the quasiparticle and quasihole
excitations. Note, however, that 
since the Hamiltonian of the NCCS theory (as well as the matrix model)
vanishes, all the allowed states are degenerate in energy.
In particular, the quasiparticle and quasihole excitations are massless
in this context. It therefore remains unclear how to obtain
a non-vanishing mass gap.
One possibility is that additional interactions must be added to 
the action.
In fact, the string construction gives a definite prescription 
for this;
if we  set $\epsilon=\delta=0$ in (\ref{newlimit}), we
recover the ordinary Seiberg-Witten limit, in which the theory
has a finite YM term, as well as coupled scalars and fermions.
It would be interesting to study this situation.

It would also be interesting if the string realization
gives some insights into the transition to 
a Wigner crystal phase at low filling fraction.
The relevant physical gauge-invariant observables
for this problem are correlation functions
of the D-particle density $J_{D0} (p)$.
It was suggested in \cite{susskind}
to calculate the correlation functions
by mapping to the noncommutative gauge theory
in a power series expansion in $\theta$.
We now have an exact expression
for the Seiberg-Witten map
\cite{okawa_ooguri2,mukhi_suryanarayana,liu_michelson,liu},
and the D-particle density $J_{D0} (p)$
is given by the open Wilson line (\ref{wilson_line}).
It would therefore be important
to study correlation functions of
the open Wilson line in NCCS theory.

Another open problem is how to describe the hierarchy of different
filling fractions in the brane picture. 
This hierarchy was explained in \cite{Haldane} by an iterative
procedure of quantizing a gas of quasiparticles or quasiholes 
of a fluid with a given filling fraction, to produce a fluid 
with a different filling fraction. The result is
\be
 \nu =
 \frac{\strut 1}
 {\displaystyle k + \frac{\strut \alpha_1}
  {\displaystyle p_1 + \frac{\strut \alpha_2}
    {\displaystyle p_2 + {}_{\ddots}
      }}}
\ee
where $k=1,3,5,\ldots$, $p_i = 2, 4, 6,\ldots$, and 
$\alpha_i = \pm 1$, depending on whether one uses quasiparticles 
or quasiholes.
For example, the $\nu = 2/3$
state is obtained from the $\nu = 1$ state 
by choosing $\alpha_1 = 1$ and $p_1 =2$, and 
the $\nu = 2/5$ state is obtained from the $\nu = 1/3$ state
with $\alpha_1=-1$ and $p_1=2$.

We have seen that a filling fraction $n/k$
fluid can be described by a configuration of $n$ D2-branes
and $2k$ D8-branes in a background $B$ field.
Let us propose a possible scenario
which accommodates the above hierarchy structure.
Consider the case with $\nu = 2/5$
by starting with $\nu = 1/3$.
Since $1/3=5/15$, we can also realize $\nu = 1/3$
by $2 \times 15$ D8-branes and $5$ D2-branes.
The world-volume theory is $U(5)$ NCCS theory
at level $k=15$.
If we increase the filling fraction by
decreasing $G_2$, while keeping the total D-particle charge fixed,
quasiparticles will appear.
Since quasiparticles correspond to the ends of strings on the
D2-branes, each carries a charge which is $1/15$ the unit D-particle
charge. However, in order to make a $U(5)$ invariant state one must
combine five strings ending on the five different D2-branes.
This $U(5)$ singlet quasiparticle then actually has a charge
$5/15 = 1/3$. This is the true quasiparticle of the $\nu=1/3$
state. As we continue to decrease $G_2$ more and more quasiparticles
are produced, until the filling fraction reaches the value $2/5$.
Here we notice that
\begin{equation}
\frac{2}{5} = \frac{1}{3} + \frac{1}{15}
= \frac{5}{15} + \frac{1}{15} = \frac{6}{15} \;.
\end{equation}
We therefore propose that at this point the quasiparticles
condense, much like the original D-particles, to form another D2-brane,
giving a total of $n=6$, and thus $\nu = 6/15 = 2/5$.
It is straightforward to extend this scenario
down the hierarchy by starting with more D8-branes.
To argue that this is more than an amusing scenario, however,
requires a better understanding
of the dynamics of quasiparticles and quasiholes.
In the matrix-model picture, construction of multi-centered
solutions of quasiparticles and quasiholes,
and the identification of their collective coordinates
would also be important.

\section*{Acknowledgments}
Y.O. would like to thank
Koji Hashimoto, Shinji Hirano, Ikuo Ichinose and Hirosi Ooguri
for useful discussions.
J.B. would like to thank Simeon Hellerman,
Lenny Susskind, and Mark van Raamsdonk
for helpful discussions.
The work of O.B. was supported in part by the DOE under
grant no. DE-FG03-92-ER40701, and by a Sherman Fairchild
Prize Fellowship.
The research of Y.O. was supported in part by
the DOE grant DE-FG03-92ER40701.
The work of J.B. was supported in part by
the DOE grant DE-AC03-76SF00515.

\end{document}